\documentclass[twocolumn,showpacs,preprintnumbers,nofootinbib,amsmath,amssymb]{revtex4}

\usepackage{graphicx}
\usepackage{amsfonts,latexsym}
\usepackage{euscript,amssymb}
\usepackage{epsfig}
\usepackage{dcolumn}
\usepackage{bm}

\catcode`\@=11
\def\Let@{\relax\iffalse{\fi\let\\=\cr\iffalse}\fi}
\def\vspace@{\def\vspace##1{\crcr\noalign{\vskip##1\relax}}}
\def\multilimits@{\bgroup\vspace@\Let@
 \baselineskip\fontdimen10 \scriptfont\tw@
 \advance\baselineskip\fontdimen12 \scriptfont\tw@
 \lineskip\thr@@\fontdimen8 \scriptfont\thr@@
 \lineskiplimit\lineskip
 \vbox\bgroup\ialign\bgroup\hfil$\m@th\scriptstyle{##}$\hfil\crcr}

\def\Sb{_\multilimits@}
\def\endSb{\crcr\egroup\egroup\egroup}
\def\Sp{^\multilimits@}

\newcommand{\nn}{\nonumber}
\newcommand{\be}[1]{\begin{equation}\label{#1}}
\newcommand{\ee}{\end{equation}}
\newcommand{\ba}[1]{\begin{eqnarray}\label{#1}}
\newcommand{\ea}{\end{eqnarray}}
\newcommand{\rf}[1]{(\ref{#1})}

\renewcommand{\theequation}{\arabic{section}.\arabic{equation}}

\begin{document}

\title{Sp-brane accelerating cosmologies}


\author{Viktor Baukh}\email{bauch_vGR@ukr.net} \
\author{Alexander Zhuk} \email{zhuk@paco.net}\\[2ex]
\altaffiliation[Also at ]{Department of Advanced Mathematics,
Odessa National Academy of Telecommunication, 1 Kuznechnaya St.,
Odessa 65069, Ukraine}
\affiliation{Department of Theoretical Physics, Odessa National University,\\
2 Dvoryanskaya St.,
Odessa 65026, Ukraine} \\


\date{\today}


\begin{abstract}We investigate time dependent solutions (S-brane solutions)
for product manifolds consisting of factor spaces where only one
of them is non-Ricci-flat. Our model contains minimally coupled
free scalar field as a matter source. We discuss a possibility of
generating late time acceleration of the Universe. The analysis is
performed in conformally related Brans-Dicke and Einstein frames.
Dynamical behavior of our Universe is described by its scale
factor. Since the scale factors of our Universe are described by
different variables in both frames, they can have different
dynamics.
Indeed, we show that with our S-brane ansatz in the Brans-Dicke
frame the stages of accelerating expansion exist for all types of
the external space (flat, spherical and hyperbolic). However,
applying the same ansatz for the metric in the Einstein frame, we
find that a model with flat external space and hyperbolic
compactification of the internal space is the only one with the
stage of the accelerating expansion. Scalar field can prevent this
acceleration. It is shown that the case of hyperbolic external
space in Brans-Dicke frame is the only model which can satisfy
experimental bounds for the fine structure constant variations. We
obtain a class of models where a pare of dynamical internal spaces
have fixed total volume. It results in fixed fine structure
constant. However, these models are unstable and external space is
non-accelerating.
\end{abstract}

 \maketitle


\section{\label{sec:1}Introduction}

\setcounter{equation}{0}

Recent astronomical observations abundantly evidence that our
Universe underwent stages of accelerating expansion during its
evolution. There are at least two of such stages: early inflation
and late time acceleration. The latter began approximately at the
redshift $z \sim 1 $ and continues until now. Thus, the
construction and investigation of models with stages of
acceleration is one of the main challenge of the modern cosmology.

Among such models, the models originated from fundamental theories
(e.g. string/M-theory) are of the most of interest. For example,
it was shown that some of spacelike brane (S-brane) solutions have
a stage of the accelerating expansion.
We remind that in
$D$-dimensional manifold S$p$-branes are time dependent solutions
with ($p+1$)-dimensional Euclidean world-volume and apart from
time they have $(D-p-2)$-dimensional hyperbolic, flat or spherical
spaces as transverse/additional dimensions \cite{GS}:
\ba{0.1} \nonumber ds_D^2 = -e^{2\gamma(\tau )} d\tau^2 &+&
a_0^2(\tau )(dx_1^2 + \ldots + dx_{p+1}^2)\\ &+ &a^2_1(\tau )
d\Sigma^2_{(D-p-2),\, \sigma}\, \, , \ea
where $\gamma (\tau )$ fixes the gauge of time, $a_0 (\tau )$ and
$a_1 (\tau )$ are time dependent scale factors and $\sigma =
-1,0,+1 $ for hyperbolic, flat or spherical spaces
respectively\footnote{Slightly generalized ansatz where the
$(D-p-2)$-dimensional transverse space consists of the
$k$-dimensional hyperspace $\Sigma_{k,\sigma}$ and
$(q-k)$-dimensional Euclidean space was considered in \cite{CGG}.
Here, $D-p-2 = k+q$.}. Obviously, $p=2$ if brane describes our
3-dimensional space. These branes usually known as SM2-branes if
original theory is 11-dimensional M-theory and SD2-branes in the
case of 10-dimensional Dirichlet strings. For this choice of $p$,
the evolution of our Universe is described by the scale factor
$a_0$. In general, the scale factor $a_1$ can also determine the
behavior of our 3-dimensional Universe. Hence, $D-p-2=3$ and we
arrive to SM6-brane in the case of the M-theory and SD5-brane for
the Dirichlet string. Usually, S$p$-brane models include form
fields (fluxes) and massless scalar fields (dilatons) as a matter
sources. If SD$p$-branes are obtained by dimensional reduction of
11-dimensional M-theory, then the dilaton is associated with the
scale factor of a compactified 11-th dimension.

Starting from \cite{GS}, the S-brane solutions were also found,
e.g., in Refs. \cite{CGG,Iv2,Oh1,Burgess}. It was quite naturel to
test these models for the accelerating expansion of our Universe.
Really, it was shown in \cite{Oh2} that the SM2-brane as well as
the SD2-brane have stages of the accelerating behavior. This
result generalizes conclusions of \cite{TW} for models with
hyperbolic compact internal spaces. Here, the cosmic acceleration
(in Einstein frame) is possible due to a negative curvature of the
internal space that gives a positive contribution into an
effective potential. This acceleration is not eternal but has a
short period and the mechanism of such short acceleration was
explained in \cite{EG}. It was indicated in \cite{Oh2} that the
solution of \cite{TW} is the vacuum case (the zero flux limit) of
the S-branes. It was natural to suppose that if the acceleration
takes place in the vacuum case, it may also happen in the presence
of fluxes. Indeed, it was confirmed for the case of the compact
hyperbolic internal space. Even more, it was found that periods of
the acceleration occur in the cases of flat and spherical internal
spaces due to the positive contributions of fluxes into the
effective potential.

It is worth of noting that accelerating multidimensional
cosmological models are widely investigated for last few years for
different types of models. In general, such models can be divided
into two main classes\footnote{Apart of these models, interesting
accelerating cosmologies following from non-linear models were
proposed in \cite{NO}}. First class consists of models where the
internal spaces are stabilized and the acceleration is achieved
due to a positive minimum of an effective potential which plays
the role of a positive cosmological constant. General method for
stabilization of the internal spaces was proposed in \cite{GZ1}
and numerous references can be found, e.g., in Refs.
\cite{GSZ,GMZ,Dark}. Models where both external (our) space and
internal spaces undergo dynamical behavior constitute the second
class of models. These models were considered, e.g., in
\cite{Gu,CV,Mohammedi} where a perfect fluid plays the role of a
matter source and the cosmic acceleration happens in Brans-Dicke
frame. Obviously, the S-branes accelerating solutions belong to
the second class of models. Along with mentioned above Ref.
\cite{Oh2}, the accelerating S-brane cosmologies (in Einstein
frame) were obtained and investigated, e.g., in Refs.
\cite{Roy,W2,GKL,CHNOW,Neupane}. Closely related to them
accelerating solutions were also found in Refs. \cite{CHNW,Iv3}
(see also general discussion on inflationary cosmologies with the
sum of exponential effective potentials in \cite{W3}; the complete
classification of solutions for such models according to their
late-time behavior is given in \cite{Jarv}). It should be noted
that some of these solutions are not new ones but either
rediscovered or written in different parametrization (see
corresponding comments in Refs. \cite{Iv2,Iv3}). For example, the
first vacuum solution for a product manifold (consisting of
$(n-1)$ Ricci-flat spaces and one Einstein space with non-zero
constant curvature) was found in \cite{Iv1}\footnote{The first
quantum solutions as well as the euclidian classical solution for
this model in the presence of a massless minimally coupled scalar
field were obtained in \cite{Zhuk1}.}. This solution was
generalized to the case of a massless scalar field in Refs.
\cite{BZpositive,BZnegative}. Obviously, solutions in Refs.
\cite{Iv1,BZpositive,BZnegative} are the zero flux limit of the
Sp-branes and the result of \cite{TW} is a particular case of
\cite{BZnegative}\footnote{However, the period of accelerating
expansion was not singled out in \cite{BZnegative}. }. Some of
solutions in \cite{CHNOW,CHNW} coincides with corresponding
solutions in Refs. \cite{Iv1,BZpositive,BZnegative,BZ3}. An
elegant minisuperspace approach for the investigation of the
product space manifolds consisting of Einstein spaces was proposed
in \cite{IMZ}. Here, it was shown that the equations of motion
have the most simple form in a harmonic time gauge\footnote{For
Eq. \rf{0.1}, it  reads $\gamma = (p+1)\ln a_0 + (D-p-2)\ln a_1 $.
In the harmonic time gauge, time satisfies equation $ \Delta [g]
\tau = 0$ \cite{IMZ}.} because the minisuperspace metric is flat
in this gauge. Even if the authors of the mentioned above papers
did not aware of it, they intuitively used this gauge to get exact
solutions. New solutions can be also generated (from the known
solutions) with the help of a topological splitting when Einstein
space with non-zero curvature is splitted into a number of
Einstein spaces of the same sign of the curvature (see Refs.
\cite{GIM,split}). This kind of solutions was found, e.g., in
Refs. \cite{CHNOW,CHNW}.

Our paper is devoted to a model with the product of $n$ Einstein
spaces where all of them are Ricci-flat but one with positive or
negative curvature. We include massless scalar field as a matter
source. As we mentioned above, the general solutions for this
model was found in our papers \cite{BZpositive,BZnegative}. Here,
all factor spaces are time dependent. Obviously, these solutions
are the zero flux limit of the S$p$-branes. The aim of the present
investigations is twofold.

First, we give the detail analysis for the accelerating behavior
of the external (our) space. At this stage, both the Ricci-flat
space and non-zero curvature space may play the role of our
Universe. The investigation is conducting in Einstein as well as
Brans-Dicke frames.
The transition between these two frames is performed with the help
of the conformal transformation of the metric of the external
spacetime. Such transformation does not destroy neither
factorizable structure of $D-$dimensional metric ansatz nor the
topology of factor spaces. However, scale factors of our Universe
are described by different variables in the Brans-Dicke and
Einstein frames. These variables are connected with each other via
conformal transformation (see Appendix). Moreover, synchronous
times are also different in both frames. Obviously, these
different scale factors may behave differently with corresponding
synchronous times. Precisely this interpretation we bear in mind
when we write about different behavior of our Universe in
different frames. For example we show that in Brans-Dicke frame,
stages of the accelerating expansion exist for all types of the
external space (flat, spherical and hyperbolic). However, in
Einstein frame, the model with flat external space and hyperbolic
compactification of the internal space is the only one with the
stage of the accelerating expansion, in agreement with the results
Refs. \cite{TW,CHNOW}. A new result here is that scalar field
can prevent the acceleration in the Einstein frame.

Second, we investigate the variation of the fine structure
constant in our model. It is well known that dynamical internal
spaces result in the variations of the fundamental constants (see,
e.g., Refs. \cite{GSZ,CV} and references therein). For example,
the fine structure constant is inversely proportional to the
volume of the internal space.
However, there are strong experimental restrictions for the
variations of the fundamental constants (see, e.g., \cite{Uzan}).
Thus, any multidimensional cosmological models with time dependent
internal spaces should be tested from this point of view. In our
paper, we show that considered model have a significant problem to
satisfy these limitations for the variation of the fine structure
constant. The case of the hyperbolic external space in the
Brans-Dicke frame is the only possibility to avoid this problem,
if there is no other way to explain the constancy of the effective
four-dimensional fundamental constants in multidimensional models.
For example, we propose models with the hyperbolic or spherical
external space and two Ricci-flat internal spaces where the total
volume of the internal spaces is the constant. Here, the dynamical
factors of the internal spaces mutually cancel each other  in the
total volume element. Thus, the effective fundamental constants
remain really constant in spite of the dynamical behavior of the
internal spaces. However, this model is unstable and the external
space is non-accelerating. Anyway, such models are of special
interest because indicate a possible way to avoid the fundamental
constant variations in higher-dimensional theories.

The paper is structured as follows. In section \ref{sec:2}, we
explain the general setup of our model and present the exact
solutions for a product manifold consisting of two factor spaces
where only one of them is non-Ricci-flat. These solutions is
carefully investigated in section \ref{sec:3} (spherical factor
space) and \ref{sec:4} (hyperbolic factor space) for the purpose
of the accelerating behavior of the external space. In section
\ref{sec:5}, we compare the rate of variations of the fine
structure constant in our accelerating models with the
experimental bounds. In section \ref{sec:6}, we obtain and discuss
a solution with three factor spaces where two dynamical internal
spaces have the fixed total volume. The main results are
summarized in the concluding section \ref{sec:7}.


\section{\label{sec:2}The model and solutions}

\setcounter{equation}{0}

In this section we present our model and give a sketchy outline of
the derivation of exact solutions. A more detailed description can
be found in our papers \cite{Zhuk1,BZpositive,BZnegative}.

We consider a cosmological model with a slightly generalized
metric \rf{0.1} in the form
\be{1.1}
g = -e^{2\gamma (\tau )}d\tau\otimes d\tau +
\sum_{i=0}^{n-1}e^{2\beta^i (\tau )}g^{(i)}\, ,
\ee
which is defined on a multidimensional manifold $M$ with product
topology
\be{1.2}
M=\mathbb{R}\times M_0\times\ldots\times M_{n-1}\, .
\ee
Let manifolds $M_i$ be $d_i$-dimensional Einstein spaces with
metric $g^{(i)}$, i.e.
\be{1.3}
R_{m_in_i}[g^{(i)}] = \lambda^i g^{(i)}_{m_in_i}\, , \quad m_i,n_i
= 1, \ldots ,d_i
\ee
and

\be{1.4}
R[g^{(i)}]=\lambda^id_i \equiv R_i\, .
\ee
In the case of constant curvature spaces parameters $\lambda^i$
are normalized as $\lambda^i = k_i(d_i-1)$ with $k_i=\pm 1,0$.

With total dimension $D=1+\sum_{i=0}^{n-1}d_i$, $\kappa_D^2$ a
D-dimensional gravitational constant, $\varphi$ a massless
minimally coupled scalar field, and $S_{YGH}$ the standard
York-Gibbons-Hawking boundary term, we consider an action of the
form
\be{1.5}
S = \frac{1}{2\kappa ^2_D}\int\limits_M {d^Dx} \sqrt {\vert g\vert
} \left( R[g] -g^{MN} \partial_M\varphi\partial_N\varphi\right) +
S_{YGH} \, .
\ee

This action encompasses the truncated bosonic sectors of various
supergravity theories. For example, for $D=11$ and in the absence
of scalar field, it represents the low energy limit of the
M-theory, and for $D=10$, it relates to the 10-dimensional
supergravity. However, for generality, we perform the analysis
with arbitrary $D$ in the presence of scalar field, specifying the
value of $D$ only for illustration of particular examples. For our
cosmological model, scalar field is homogeneous and depends only
on time.

We restrict our consideration to the case when only one of the
spaces $M_i$ is not Ricci-flat: $R_0 \ne 0,\, R_i =0, \,
i=1,\ldots ,n-1.$ Taking into account the homogeneity of our
model, the action $S$ is reduced to the form:
\ba{1.6} S &=& \mu \int {Ld\tau } \\ \nonumber &=& \mu \int d\tau
\left\{ \frac{1}{2}e^{ - \gamma + \gamma _0 }[G_{ij} \dot {\beta
}^i\dot {\beta }^j + \dot {\varphi }^2] - e^{\gamma - \gamma _0
}U\right\} \, , \ea
where
\be{1.7}
 U = - \frac{1}{2}e^{2\gamma _0 }R_0 e^{ - 2\beta ^0}
 \ee
 is the
 potential, $\gamma_0 = \sum_{i=0}^{n-1}d_i\beta^i$, $G_{ij}=d_i
 \delta _{ij}-d_id_j\quad (i,j = 0,\ldots
 ,n-1)\quad $is the minisuperspace metric, $\mu =
 \prod_{i=0}^{n-1}V_i/\kappa^2$, and $V_i = \int_{M_i } {d^{d_i
}y(\det (g_{m_i n_i }^{(i)} )^{1/2}} )$ is the volume of $M_i$
(modulo the scale factor of the internal space).

It can be easily seen that the Euler-Lagrange equations for
Lagrangian \rf{1.6} as well as the constraint equation $\partial
L/\partial \gamma =0 $ have the most simple form in the harmonic
time gauge $\gamma = \gamma_0 = \sum_{i=0}^{n-1}d_i\beta^i$
\cite{IMZ}. The corresponding solutions can be found in
\cite{Zhuk1,BZpositive,BZnegative}. For simplicity we consider a
model with two factor spaces ($n=2$). All our conclusions can be
easily generalized to a model with $n>2$ factor spaces. For two
component cosmological model the explicit expressions for the
scale factors and scalar field as functions of harmonic time read:
\ba{1.8}
 a_0 (\tau ) &=& \exp (\beta^0(\tau )) =a_{(c)0}
 \exp ( - \frac{\xi _1 }{d_0 - 1}\tau )
 \times\frac{1}{g_\pm
(\tau )}\nonumber \, ,\\
 a_1 (\tau ) &=& \exp( \beta^1(\tau )) =A_1
 \exp (\frac{\xi _1 }{d_1 }\tau )\, , \\
  \varphi (\tau ) &=& p^2\tau + q\nonumber\, ,
\ea
where
\be{1.9}  g_ + = \cosh^{1/(d_0 - 1)}\left( {\xi _2 \tau }
\right),\,  \quad (-\infty < \tau < +\infty ), \ee for $R_0
> 0\,$ and
\be{1.10}  g_ - = \sinh^{1/(d_0 - 1)}\left( {\xi _2 |\tau |}
\right),\, \quad (|\tau |>0), \ee for $R_0 <
 0$.
 Here, $a_{(c)0}=A_0(2\varepsilon/|R_0|)^{1/2(d_0-1)},\, \xi_1
= [d_1(d_0-1)/(D-2)]^{1/2} p^1,\, \xi _2=[(d_0 -
1)/d_0]^{1/2}(2\varepsilon )^{1/2}$ and
$2\varepsilon=(p^1)^2+(p^2)^2$. Parameters $A_0 , A_1, p^1, p^2$
and $q$ are the constants of integration and $A_0, A_1$ satisfy
the following constraint: $A_0^{d_0}A_1^{d_1}=A_0$. It was shown
in \cite{Zhuk1} that $p^1$ and $p^2$ are the momenta in the
minisuperspace ($p^1$ is related to the momenta of the scale
factors and $p^2$ is responsible for the momentum of scalar field)
and $\varepsilon$ plays the role of energy.

In what follows, we consider the case of positive $\varepsilon$
and without loss of generality we chose $2\varepsilon = 1\,
\Rightarrow \, (p^1)^2+(p^2)^2=1$. We also put $q=0$. It is also
convenient to consider the dimensionless analogs of the scale
factors: $a_0(\tau)\rightarrow a_0(\tau )/a_{(c)0}$ and $ a_1(\tau
) \rightarrow a_1(\tau)/A_1$. This choice does not affect the
results but simplifies the analysis. So, below we investigate
these dimensionless scale factors denoting them by the same
letters as the dimensional scale factors.

The solution \rf{1.8} is written in the harmonic time gauge. The
synchronous time gauge (in other words, the proper time gauge)
corresponds to $\gamma =0$. This choice takes place in Brans-Dicke
frame. In Einstein frame the synchronous gauge is different. The
relation between these gauges in different frames is presented in
Appendix and it depends on the choice of the external and internal
spaces.  In our analysis both $M_0$ and $M_1$ can play the role of
the external and internal spaces.

The dynamical behavior of the factor spaces is characterized by
the Hubble parameter
\be{1.11}
H_i(t) = \frac{\dot a_i(t)}{a_i(t)}\, , \quad i=0,1
\ee
and the deceleration parameter
\be{1.12}
q_i(t)=-\frac{\ddot a_i(t)}{a_i(t)}\, , \quad i=0,1\, ,
\ee
where the overdots denote the differentiation with respect to the
synchronous time $t$ which is connected with the harmonic time
$\tau$ as follows:
\be{1.13}
dt=f(\tau )d\tau \quad \Longrightarrow \quad t(\tau ) =
\int_{-\infty}^{\tau} f(\tau )d\tau\, ,
\ee
where the function $f(\tau )$ is defined in accordance with Eqs.
\rf{a.6} and \rf{a.7} and we fix the constant of integration in
such a way that $t \rightarrow 0$ for $\tau \rightarrow -\infty$.
In Eqs. \rf{1.11} - \rf{1.13}, the quantities $a_i$ and $t$ are
related to both the Brans-Dicke and the Einstein frames and the
exact form of $f(\tau )$ depends on the choice of the frame (in
the Einstein frame it depends also on the choice of the external
space). Since in our model $f(\tau )>0$, the synchronous time
$t(\tau )$ is a monotone increasing function of the harmonic time.
The expressions for the parameters $H_i$ and $q_i$ can be
rewritten with respect to the harmonic time:
\be{1.14}
H_i (t(\tau )) = \frac{1}{a_i }\frac{d}{dt}a_i = \frac{1}{f(\tau
)a_i
(\tau )}\frac{da_i (\tau )}{d\tau }
\ee
and
\ba{1.15} \nonumber - q_i (t(\tau )) &=& \frac{1}{f^2(\tau )a_i
(\tau )}\Biglb(\frac{d^2a_i (\tau )}{d\tau ^2}\\ &-&
\frac{1}{f(\tau )}\frac{df(\tau )}{d\tau }\frac{da_i (\tau
)}{d\tau }\Bigrb)\, . \ea
With the help of these equations we can get a qualitative picture
of the dynamical behavior of the factor spaces in synchronous time
via the solutions \rf{1.8} in the harmonic time gauge. More
detailed information can be found from the exact expressions for
$a_i(t)$. To get it, we should calculate the integral \rf{1.13}
which provides the connection between harmonic and synchronous
times.
However, the function $f(\tau )$ is a transcendental function and
the integral \rf{1.13} is not expressed in elementary functions.
Hence, we shall
analyze equations \rf{1.14}, \rf{1.15} and asymptotic expressions
for $a_i(t)$ to get an information about the dynamics of the
factor spaces in synchronous time.
To confirm our conclusions graphically, we shall use the
$\it{Mathematica\, 5.0}$ to draw the dynamical behavior of
$a_i(t)$ for full range of time $t$ (for a particular choice of
parameters of the model).

\section{\label{sec:3}Spherical factor space}

\setcounter{equation}{0}

In this section we investigate models where the factor space $M_0$
has the positive curvature $R_0>0$. We split our consideration
into two separate subsections where calculation will be done in
Brans-Dicke and Einstein frames correspondingly.

\subsection{\label{sec:3.1}Brans-Dicke frame}

 In the case of spherical space $M_0$ the scale factors
\rf{1.8} have the following asymptotic forms:
\ba{2.1}
 a_0 (\tau )\vert _{\tau \to \pm \infty } &\simeq& 2^{\frac{1}{d_0 -
 1}}\exp{\biglb(
 { - \frac{\xi _1 \pm \xi _2 }{d_0 - 1}\tau }\bigrb)}\, , \\
 \label{2.2} a_1 (\tau )\vert _{\tau \to \pm \infty } &=& \exp{\biglb(\frac{\xi _1 }
 {d_1 }\tau \bigrb)}\, ,
\ea
where we use the condition $\xi_2>0$. It can be easily seen that
the asymptotic behavior depends on signs of $\xi_1 \pm \xi_2$ and
$\xi_1$.

The comparison of Eqs. \rf{1.13} and \rf{a.6} gives the expression
for the function $f(\tau )$:
\ba{2.3} f(\tau ) &=& f_{+BD} (\tau ) = e^{\gamma _0 } \\ &=&
a_0^{d_0 } a_1^{d_1 }=  \frac{\exp\biglb( - \frac{\xi _1 }{d_0 -
1}\tau \bigrb)}{ g_ + ^{ d_0 } (\tau )},\,  \quad \tau \in
(-\infty , +\infty) \nonumber\ea
with the asymptotes
\be{2.4} f_{+BD} (\tau )\vert _{\tau \to \pm \infty }\,  \simeq \,
2^{\frac{d_0 }{d_0 - 1}}\exp \biglb( - \frac{\xi _1 \pm d_0 \xi _2
}{d_0 - 1}\tau \bigrb). \ee
Thus, from Eq. \rf{1.13} we obtain the asymptotic expression for
the synchronous time
\be{2.5} t - t_0 \vert _{\tau \to +\infty } \simeq - 2^{\frac{d_0
}{d_0 - 1}}\frac{d_0 - 1}{\xi _1 + d_0 \xi _2 }e^{ { - \frac{\xi
_1 + d_0 \xi _2 }{d_0 - 1}\tau } }\, , \ee
which enable us to rewrite the asymptotes \rf{2.1} and \rf{2.2} in
the synchronous time gauge:
\be{2.6} a_0 (t)\vert _{t \to t_0 } \simeq
2^{\frac{1}{d_0-1}}\Biglb[ {\frac{(d_0 \xi _2 + \xi _1 )(t_0 -
t)}{2^{\frac{d_0 }{d_0 - 1}}(d_0 - 1)}} \Bigrb]^{\frac{\xi _1 +
\xi _2 }{\xi _1 + d_0 \xi _2 }}\, ,\ee~{ \be{2.7} a_1 (t)\vert _{t
\to t_0 } \simeq \Biglb[ {\frac{(d_0 \xi _2 + \xi _1 )(t_0 -
t)}{2^{\frac{d_0 }{d_0 - 1}}(d_0 - 1)}} \Bigrb]^{ - \frac{\xi _1
(d_0 - 1)}{d_1 (\xi _1 + d_0 \xi _2 )}}\, . \ee}

Additionally, it can be easily seen that conditions
\be{2.8}
- d_0\xi_2 < \xi_1 < d_0\xi_2
\ee
provide the convergence of the integral \rf{1.13} for any value of
$\tau $ from the range $(-\infty , +\infty )$. Thus, infinite
range of $\tau$ is mapped onto the finite range of $t$. We remind
also that the synchronous time $t(\tau )$ is a monotone increasing
function of the harmonic time.

Now, with the help of the expression \rf{2.3} for $f_{+BD} (\tau
)$, the Hubble and the deceleration parameters are easily obtained
from Eqs. \rf{1.14} and \rf{1.15}:
\ba{2.9} H_0 &=& - \frac{1}{f_{+BD} (\tau )}\frac{\xi _1 + \xi _2
\tanh(\xi _2 \tau )}{d_0 - 1} \, ,\\
 \label{2.10}  q_0 &=&  \frac{\xi _2}{f_{+BD}^2 (\tau )}\frac{\xi _2 + \xi _1 \tanh(\xi
_2 \tau )}{d_0 - 1}
\ea
for the factor space $M_0$ and
\ba{2.11}
 H_1 &=& \frac{1}{f_{+BD} (\tau )}\, \frac{\xi _1}{d_1} \, , \\
 \label{2.12} - q_1 &=& \frac{\xi _1}{f_{+BD}^2 (\tau )}  \\ \nonumber &\times& \frac{(D - 2)\xi _1 + d_0 d_1
\xi _2 \tanh(\xi _2 \tau ))}{d_1^2 (d_0 - 1)} \ea
for the Ricci-flat factor space $M_1$.

The following analysis depends on the choice of the external
space. Therefore, we consider two separate cases.
\subsubsection{\label{sec:3.1.1}\bf{Spherical external space (SM6 and SD5 branes)}}
As we already wrote in Introduction, solutions in this case
describe the vacuum SM6-brane if $D=11, d_0=3$ and scalar field is
absent ($p^2 = 0  \to |p^1| = 1$) and the zero flux limit of the
SD5-brane in the presence of scalar/dilaton field if $D=10, d_0=3$
and $|p^1|\leq 1$.

Since we are looking for a solution with the dynamical
compactification of the internal space $M_1$, the parameter
$\xi_1$ should be negative: $\xi_1<0 \to p^1 <0$ (see Eq.
\rf{2.11}). Then, Eqs. \rf{2.9} and \rf{2.10} show that the
accelerating expansion of the external space $M_0$ takes place for
harmonic times
\be{2.13}
\frac{\xi_2}{|\xi_1|}<\tanh(\xi_2\tau ) < \frac{|\xi_1|}{\xi_2}
\ee
that leads to inequality
\be{2.14}
|\xi_1|>\xi_2 \quad \Longrightarrow \quad \xi_1 < - \xi_2\, .
\ee
Additionally, it can be easily proven that the inequalities
\rf{2.8} are also valid for this case (the right inequality is
obvious for negative $\xi_1$ and the left inequality follows from
the condition $|p^1| \leq 1$). Therefore, the range of the
synchronous time $t$ is finite. Thus, with the help of the
inequalities \rf{2.8} and \rf{2.14} we arrive to the following
conclusions. First, from the asymptote \rf{2.1} follows that the
factor space $M_0$ expands from zero ($\tau \to -\infty$) to
infinity ($\tau \to +\infty$) and it occurs for the finite range
of the synchronous time. It is the typical Big Rip scenario. At
the same time, the internal space $M_1$ contracts from infinity to
zero (see \rf{2.2}). Second, starting from the time $
\tanh(\xi_2\tau )= \xi_2/|\xi_1|$, the acceleration never stops
lasting until the Big Rip\footnote{In the case of a pure imaginary
scalar field the parameters $\xi_1$ and $\xi_2$ can satisfy the
inequality $\xi_1+d_0\xi_2<0$ because $|p^1|>1$. Then, starting
from the time $ \tanh(\xi_2\tau )= \xi_2/|\xi_1|$ the external
space undergoes the eternal accelerating expansion. Here, the
synchronous time $t$ runs to $+\infty$.} (because
$|\tanh(\xi_2\tau )|\leq 1 \, \forall \, \tau \in (-\infty
,+\infty )$). For example, the accelerating expansion of the $M_0$
at late synchronous times can be directly observed from \rf{2.6}
because $(\xi_1 +\xi_2) <0$. The typical behavior of the scale
factors of the external ($a_0(t)$) and internal ($a_1(t)$) factor
spaces in the synchronous time gauge is illustrated in Fig. 1.
$t^{acc}_{in}$ denotes the time of the
beginning of the external space acceleration.
\begin{figure*}[htbp]

\includegraphics[width=6.5in,height=1.8in]{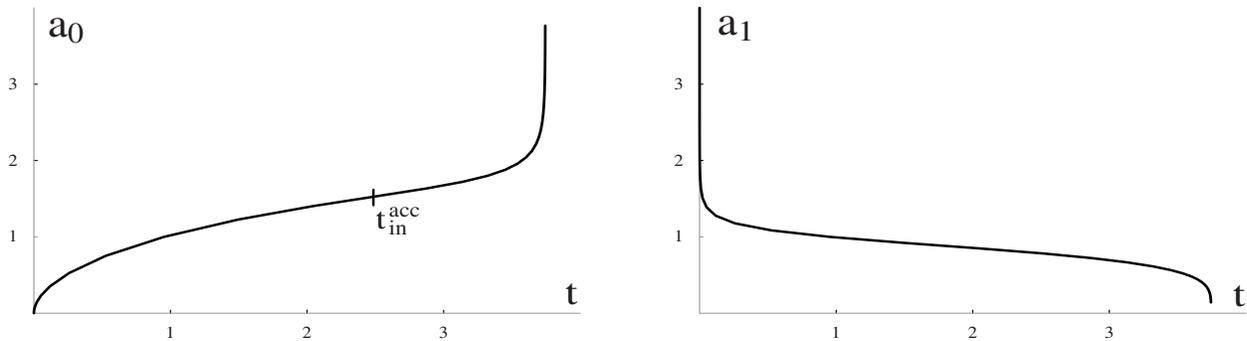}

\caption {Typical form of the external (left) and internal (right)
scale factors in Brans-Dicke frame in the case of the spherical
external space $M_0$. Specifically, it represents the zero flux
limit of the SD5-brane with $d_0=3, d_1=6$ and $p^1=-0.8$.}
\end{figure*}
\subsubsection{\label{sec:3.1.2}\bf{Ricci-flat external space (SM2 and SD2 branes)}}

Let us consider now the factor space $M_1$ as the external one.
Solutions in this case describe the vacuum SM2-brane if $D=11,
d_1=3$ and $p^2 = 0\, \, (|p^1|= 1)$  and the zero flux limit of
the SD2-brane if $D=10, d_1=3$ and $|p^1|\leq 1$.

The demand of the external space $M_1$ expansion results in the
positivity of the parameter $\xi_1$ (see Eq. \rf{2.11}): $\xi_1>0
\to 0 <p^1\leq 1$. We remind that parameter $\xi_2$ is also
positive. It is not difficult to verify that the inequalities
\rf{2.8} are also valid for the considered case. Thus, infinite
range $(-\infty ,+\infty )$ of the harmonic time $\tau $ is mapped
onto the finite range of the synchronous time $t$. According to
Eq. \rf{2.2}, for this finite synchronous time the external space
$M_1$ expands from zero value to infinity. So, we have again the
Big Rip scenario. The acceleration of the external space begins at
the time
\be{2.14a} \tanh (\xi_2\tau_a ) = -\frac{D-2}{d_0d_1} \,
\frac{\xi_1}{\xi_2} =  -\sqrt{\frac{D-2}{d_0d_1}}\, p^1 \, .\ee
Starting from this time, the acceleration of $M_1$ never stops
lasting until the Big Rip. For example, the accelerating expansion
of the $M_1$ at late synchronous times can be directly seen from
\rf{2.7} because of the negative sign of the exponent.

As it follows from the asymptote \rf{2.1}, concerning the internal
factor space $M_0$ we have two different scenarios depending on
the relation between $\xi_1$ and $\xi_2$:

1. $\xi_1 > \xi_2\, \, \Rightarrow \, \,
\sqrt{\frac{D-2}{d_0d_1}}<p^1 \leq 1$.

Here, the internal space contracts from plus infinity to zero for
a finite synchronous time. This scenario is realized e.g. for the
case of the absence of scalar field: $p^1=1$ (see Fig.2, firm
lines).

2. $0<\xi_1 \leq \xi_2\, \, \Rightarrow \, \, 0<p^1 \leq
\sqrt{\frac{D-2}{d_0d_1}}$.

In this case, the internal scale factor $a_0$ begins to expand
either from zero value for $\xi_1 < \xi_2$ or from the finite
value $2^{1/(d_0-1)}$ for $\xi_1 = \xi_2$ until its turning point
at a maximum (at the time $\tanh (\xi_2\tau )= - \xi_1/\xi_2$ (see
Eq. \rf{2.9})) and then contracts to zero value (see Fig. 2, dash
lines). Obviously, this scenario take place in the presence of
scalar field because $p^1<1$.

\begin{figure*}[htbp]

    \includegraphics[width=6.5in,height=1.8in]{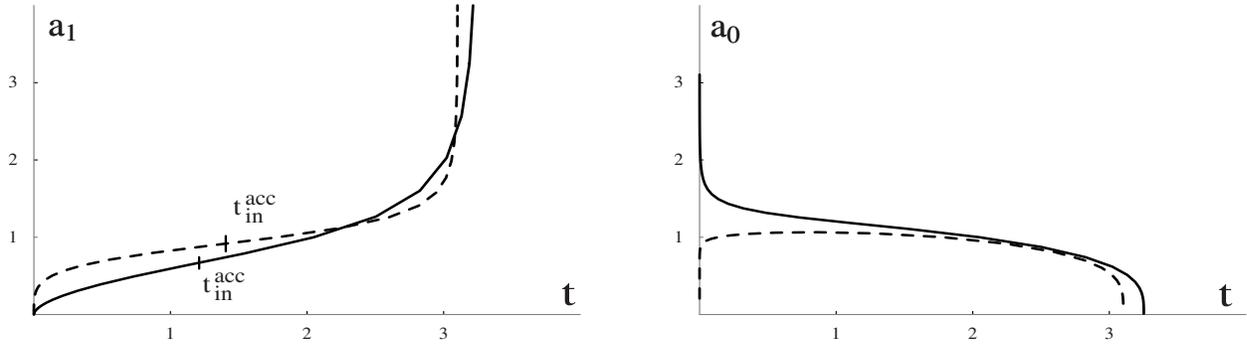}

\caption{Typical form of the external (left) and internal (right)
scale factors in Brans-Dicke frame for the Ricci-flat external
space in the cases $\xi_1 > \xi_2$ (firm lines) and $\xi_1 \leq
\xi_2$ (dash lines). Specifically, it represents the vacuum limit
of the SM2-brane with $d_0=7, d_1=3$ and $p^1=1$ (firm lines) and
the zero flux limit of the SD2-brane with $d_0=6, d_1=3$ and
$p^1=0.5$ (dash lines).}
\end{figure*}

\subsection{\label{sec:3.2}Einstein frame}
Now, we investigate the dynamical behavior of the corresponding
S$p$-branes in the Einstein frame. Similar to the Brans-Dicke
frame case, we perform our consideration for two separate cases
depending on the choice of the external factor space.

\subsubsection{\label{sec:3.2.1}\bf{Spherical external space (SM6 and SD5 branes)}}

In this case the conformal factor reads (see Eq. \rf{a.2})
\be{2.15}
\Omega = a_1^{ - \frac{d_1 }{d_0-1}} = \exp ( - \frac{\xi _1
}{d_0-1}\tau )\, .
\ee
Making use of Eqs. \rf{a.5} and \rf{a.7}, we obtain the function
$f(\tau )$
\be{2.16}
f(\tau ) = f_{+E(0)} (\tau ) = \Omega ^{ - 1} e^{\gamma _0 } =
\left[ {\cosh(\xi _2 \tau )} \right]^{-\frac{d_0}{d_0-1}}
\ee
and the scale factor of the external space
\be{2.17}
\tilde {a}_0 (\tau ) = \Omega ^{ - 1}a_{0}= g_+^{-1} =  \left[
\cosh(\xi _2 \tau )\right]^{-\frac{1}{d_0-1}}\, .
\ee
Substituting these expressions in Eqs. \rf{1.14} and \rf{1.15}, we
obtain the Hubble and the deceleration parameters
\ba{2.18}
\tilde {H}_0 (\tau ) &=& - \frac{\xi _2}{(d_0-1)f_{+E(0)} (\tau )}
\tanh(\xi
_2 \tau )\, ,\\
\label{2.19}  \tilde {q}_0 (\tau ) &=&  \frac{\xi _2^2 }{(d_0-1)
f_{_{+E(0)} }^2 (\tau )}\, .
\ea
Eqs. \rf{2.18} and \rf{2.19} clearly show that $\tilde {H}_0 (\tau
)<0$ for positive $\tau$ and $\tilde {q}_0 (\tau )>0\, \, \forall
\, \, \tau \in (-\infty ,+\infty ) $. Therefore, the external
factor space $M_0$ contracts at late times and never has the stage
of the acceleration. Obviously, this model contradicts the
observations. Here, SM6-brane corresponds to the choice of $d_1=7$
and for the SD5-brane we should take $d_1=6$.

\subsubsection{\label{sec:3.2.2}\bf{Ricci-flat external space (SM2 and SD2 branes)}}
Let the factor space $M_1$ be the external space. In this case the
conformal factor is
\be{2.20} \Omega (\tau ) = a_0^{ - \frac{d_0 }{d_1-1}} = e^ {
{\frac{d_0 }{(d_0 - 1)(d_1-1)}\xi _1 \tau } }\biglb[ {\cosh(\xi _2
\tau )} \bigrb]^{\frac{d_0 }{(d_0 - 1)(d_1-1)}}\, \ee
and for the function $f(\tau )$ and the external scale factor we
obtain the following expressions:
\ba{2.21} f(\tau ) &=& f_{+E(1)} (\tau ) = \Omega ^{ - 1}
e^{\gamma _0 }\\ \nonumber &=& e^{ {  \frac{1-d_0 -d_1}{(d_0 -
1)(d_1-1)}\xi _1 \tau } }\Biglb[ {\cosh\biglb( {\xi _2 \tau }
\bigrb)}
\Bigrb]^{-\frac{d_0d_1 }{(d_0 - 1)(d_1-1)}}\, ,\\
\label{2.22} \tilde {a}_1 (\tau ) &=&\Omega^{-1} a_1 \\ &=& e^{{
\frac{1- d_0 -d_1}{d_1(d_0 - 1)(d_1-1)}\xi _1 \tau } } \nonumber
\Biglb[ {\cosh\biglb( {\xi _2 \tau } \bigrb)} \Bigrb]^{-\frac{d_0
}{(d_0 - 1)(d_1-1)}}\, . \ea Thus, the Hubble and the deceleration
parameters of the external factor space $M_1$ read:
\begin{widetext}
\ba{2.23} \tilde {H}_1 (\tau ) &=& - \frac{1}{d_1(d_1-1)(d_0 -
1)f_{+E(1)} (\tau )}\Biglb( {(D -
2)\xi _1 + d_1d_0 \xi _2 \tanh(\xi _2 \tau )} \Bigrb)\, , \\
\label{2.24}   \tilde {q}_1 (\tau ) &=&  \frac{\Biglb( {\biglb[
{(D - 2)\xi _1 + d_0d_1 \xi _2 \tanh(\xi _2 \tau )} \bigrb]^2 +
d_0(d_0 - 1)d_1^2\xi_2^2\cosh^{-2}(\xi _2 \tau )}
\Bigrb)}{d_1^2(d_1-1)(d_0 - 1)^2f_{+E(1)}^2 (\tau )} \, . \ea
\end{widetext}
Therefore, the deceleration parameter $\tilde {q}_1 (\tau )>0\, \,
\forall \, \, \tau \in (-\infty ,+\infty ) $ and the external
space $M_1$ does not undergo the acceleration. Similar to the
previous case, the external space $M_1$ contracts at late times
(it follows from Eq. \rf{2.23} and the condition $|p^1|\leq 1$).
Hence, this model is also not of interest for us. For this case,
the SM2-brane corresponds to the choice of $d_0=7$ and for the
SD2-brane we should take $d_0=6$.

\section{\label{sec:4}Hyperbolic factor space}

\setcounter{equation}{0}

In this section we investigate models where the factor space $M_0$
has the negative curvature $R_0<0$. If this factor space is
treated as the internal one we suppose that $M_0$ is compact (see
e.g. \cite{KMST}). Similar to the previous section, we split our
consideration into two separate subsections where calculation will
be done in Brans-Dicke and Einstein frames correspondingly.

\subsection{\label{sec:4.1}Brans-Dicke frame}
As apparent from Eqs. \rf{1.8} and \rf{1.10}, the function
$a_0(\tau )$ is divergent at $\tau = 0$. This point divides the
range of $\tau$ into two separate parts: $(-\infty, 0]$ and
$[0,+\infty )$. We choose the interval $(-\infty, 0]$ because the
dynamical picture in both of these intervals is equivalent up to
the replacement $p^1 \rightarrow -p^1$ \cite{BZnegative}.

To begin with, let us first define the function $f(\tau )$
\ba{3.1} f(\tau ) &= &f_{-BD} (\tau ) = e^{\gamma _0 } = a_0^{d_0
} a_1^{d_1 }
\\\nonumber &= &\frac{\exp ( - \frac{\xi _1 }{d_0 - 1}\tau )}{g_ -
^{d_0 } (\left| \tau \right|)}\, ,\quad \tau \in (-\infty, 0] \ea
and its asymptotes
\be{3.2} f_{-BD} (\tau )\simeq \left\{
\begin{array}{c}
  2^{\frac{d_0}{d_0-1}}e^{\frac{1}{d_0-1}(\xi_1-d_0\xi_2)|\tau |)}\, , \\
  \vphantom{\exp } \\
  \biglb(\xi_2|\tau |\bigrb)^{-\frac{d_0}{d_0-1}}\, ,
\end{array}\right. \begin{array}{c}
  \tau \rightarrow -\infty \, , \\
  \vphantom{\exp } \\
  \tau \rightarrow -0\, .
\end{array}
\ee
The first asymptote $f_{-BD} (\tau ) \rightarrow 0$ in the limit
$\tau \rightarrow -\infty $ because
$(\xi_1-d_0\xi_2)<0$\footnote{It is obvious for negative $\xi_1$
and also true for positive $\xi_1$ because of $|p^1|\leq 1$.} and
the second asymptote $f_{-BD} (\tau ) \rightarrow +\infty$ in the
limit $\tau \rightarrow -0$. Thus, it can be easily seen that the
harmonic time interval $\tau \in (-\infty, 0]$ is mapped onto
synchronous time interval $t \in [0,+\infty)$ correspondingly.
These asymptotes give a possibility to connect the synchronous and
harmonic times in the corresponding limits. For example, at late
times we get the following relation:
\be{3.3}
\xi_2 t \, \simeq \, (d_0-1) \left(\xi_2|\tau
|\right)^{-\frac{1}{d_0-1}}\, , \quad \tau \rightarrow -0
\Rightarrow t \rightarrow +\infty\, .
\ee

It is also useful to present the asymptotes for the scale factors.
For the factor space $M_0$ we get:
\ba{3.4}
 a_0 (\tau )\vert _{\tau \to - \infty } &\simeq& 2^{\frac{1}{d_0 - 1}}\exp
\left( { \frac{\xi _1 - \xi _2 }{d_0 - 1}|\tau |} \right)\, , \\
 \label{3.5} a_0 (\tau )\vert _{\tau \to -0 } &\simeq& \left(\xi_2 |\tau|
 \right)^{-\frac{1}{d_0-1}}\,  \rightarrow \, +\infty\, .
\ea
The first asymptote demonstrates that there are two different
scenarios depending on the sign of the difference $\xi_1-\xi_2$.
If $\xi_1>\xi_2$, the factor space $M_0$ begins to contract from
plus infinity to a finite value and then to expand again to plus
infinity (see \rf{3.5}). If $\xi_1<\xi_2$, the factor space $M_0$
expands for all the time starting from zero to
infinity\footnote{In the exceptional case $\xi_1=\xi_2 := \xi$,
the scale factor $a_0$ reads $a_0(\tau ) = \left[(1-e^{-2\xi |\tau
| })/2\right]^{-1/(d_0-1)}$. This formula shows that the scale
factor starts from the finite value $(1/2)^{-1/(d_0-1)}$ and
expends to infinity.}. The substitution of \rf{3.3} into \rf{3.5}
shows that the Milne-type behavior of $M_0$ at late times is the
attractor solution\footnote{It can be easily verified that the
dimensional scale factor $a_0$ has the exact Milne asymptote: $a_0
(t)\vert _{t \to +\infty } \simeq t$ and for dimensional $a_1$ we
obtain $a_1 (\tau )\vert _{\tau \to - 0 } \, \rightarrow \, A_1$.}
(see e.g. \cite{BZnegative}):
\be{3.6} a_0 (t)\vert _{t \to +\infty } \simeq
\frac{1}{d_0-1}\xi_2 t \, , \ee
Concerning the factor space $M_1$ we have the following
asymptotes:
\ba{3.7}
a_1 (\tau )\vert _{\tau \to - \infty } &=&
\exp{(\frac{\xi_1}{d_1}\tau )}\, ,\\
\label{3.8} a_1 (\tau )\vert _{\tau \to - 0 } \, &\rightarrow &\,
1\, .
\ea
Here, we also have two scenarios depending on the sign of $\xi_1$.
If $\xi_1>0$, the factor space $M_1$ contracts from infinity with
the subsequent freezing at late times. If $\xi_1<0$, the factor
space $M_1$ expands from zero freezing again at late times. Thus,
the freezing of the factor space $M_1$ is the attractor behavior
at late times (see \cite{BZnegative}).

Let us define now the Hubble and the deceleration parameters. For
the factor spaces $M_0$ and $M_1$ we obtain respectively:
\ba{3.9}
H_0 &=& - \frac{1}{f_{-BD} (\tau )}\frac{\xi _1 + \xi _2 \coth(\xi
_2 \tau
)}{d_0 - 1}\, ,\\
 \label{3.10}  q_0 &=&  \frac{\xi _2}{f_{-BD}^2 (\tau )} \frac{\xi _2 + \xi _1 \coth(\xi
_2 \tau )}{d_0 - 1}
\ea
and
\ba{3.11}
 H_1 &=& \frac{1}{f_{-BD} (\tau )}\, \frac{\xi _1}{d_1} \, , \\
 \label{3.12} - q_1 &=& \frac{\xi _1}{f_{-BD}^2 (\tau )} \\\nonumber  &\times& \frac{(D - 2)\xi _1 + d_0 d_1
\xi _2 \coth(\xi _2 \tau )}{d_1^2 (d_0 - 1)}\, .
\ea
With the help of these expressions we can analyze the factor
spaces from the point of their acceleration. Again, the analysis
depends on the choice of the external space.

\subsubsection{\label{sec:4.1.1}\bf{Hyperbolic external space (SM6 and SD5 branes)}}

Usually, we are looking for a model with expending external space
and contracting (or static) internal one. As it follows from Eqs.
\rf{3.9} and \rf{3.11}, the choice $\xi_1\leq 0$ guarantees these
conditions. However, the external factor space is a decelerating
one at all times because $q_0>0 \, \, \forall\, \,  \tau \in
(-\infty ,0]$ (see Eq. \rf{3.10}). Therefore, in the rest of this
subsection we investigate the case of positive $\xi_1 >0\,
\rightarrow \, p^1>0$ with expending internal space. In spite of
the expending character of the internal space, Eq. \rf{3.8} shows
that this space goes asymptotically to a constant value ("freezed
out") at late times. We suppose that this value is less than the
Fermi length $L_F \sim 10^{-17}$cm. It makes the internal space
unobservable at late times.

Obviously, for positive $\xi_1$ we have two scenarios:

1. $\quad \xi_1 > \xi_2$.

Here, the external space $M_0$ after the contraction from infinity
to a finite value starts to expend at the time
\be{3.13}
\coth(\xi _2 \tau _e ) = - \frac{\xi _1 }{\xi _2 } = - \sqrt
{\frac{d_0 d_1 }{D - 2}} \, {p^1}
\ee
asymptotically approaching to the attractor $a_0 \sim t\, \,
(t\rightarrow +\infty )$. At all stages of its evolution the
factor space $M_0$ has the accelerating behavior: $q_0<0 \, \,
\forall \, \, \tau \in (-\infty ,0]$. This scenario is realized
e.g. for the case of the absence of scalar field: $p^1=1$ (see
Fig.3, firm lines, where the convex curve $a_0$ has positive
second derivative/acceleration for all $t\in [0,+\infty)$).

2. $\quad 0<\xi_1 \leq \xi_2$.

Here, the external space $M_0$ expends for all time $\tau \in
(-\infty ,0]$ stating from zero ( for $\xi_1<\xi_2$) or from a
finite value (for $\xi_1=\xi_2$) asymptotically approaching to the
attractor $a_0 \sim t\, \, (t\rightarrow +\infty )$. The
acceleration begins at the time
\be{3.14}
\coth(\xi _2 \tau _a ) = -\frac{ \xi _2 }{\xi _1 } = - \sqrt
{\frac{D-2 }{d_0 d_1}} \frac{1}{p^1}\, .
\ee
This equation is satisfied for $p^1 < \sqrt{(D-2)/d_0d_1}<1$, i.e.
in the presence of sufficiently dynamical scalar field. The
typical behavior of the scale factors in the synchronous time
gauge for this type of scenarios is illustrated in Fig. 3 (dash
lines).

It is worth of noting that to draw the graphics  in synchronous
time, we use in the integral \rf{1.13} the exact expressions for
the function $f(\tau )$ rather than its asymptotes. It can result
in a proper shift between an analytic estimate (for the late
times) and a graphical plotting. For example, corresponding shift
for the linear asymptote \rf{3.6} has the form of $a_0 (t)\vert
_{t \to +\infty } \simeq \frac{1}{d_0-1}\xi_2 (t+t_0)$ where $t_0
= \lim\limits_{\tau \to 0} \int_{-\infty}^{\tau}
[f(\eta)-\left(\xi_2|\eta |\right)^{-\frac{d_0}{d_0-1}}]d\eta$.
Because function $f(\tau )$ depends on parameter $\xi_1$, the firm
and dash lines in the left picture of Fig.3 acquire the late time
relative shift with respect to each other.

\begin{figure*}[htbp]

    \includegraphics[width=6.5in,height=1.8in]{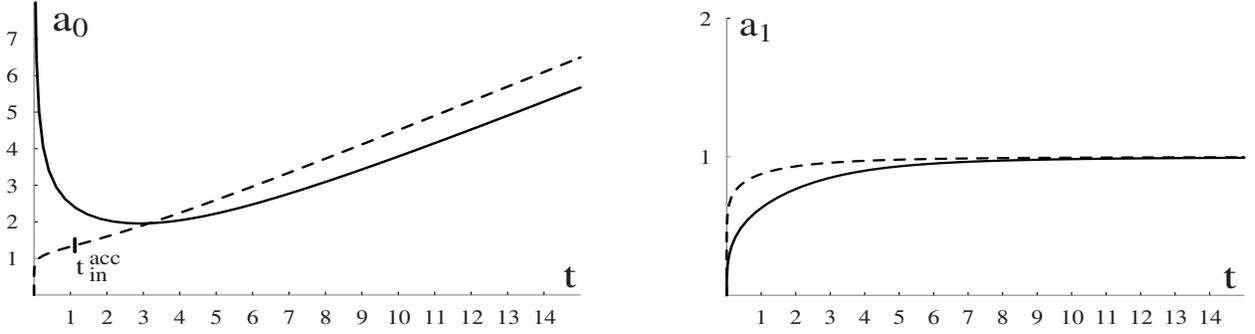}

\caption{Typical form of the external (left) and internal (right)
scale factors in Brans-Dicke frame  for the hyperbolic external
space $M_0$ in the case $\xi_1>\xi_2$ (firm lines) and
$\xi_1<\xi_2$ (dash lines). Specifically, it represents the vacuum
limit of the SM6-brane with $d_0=3, d_1=7$ and $p^1=1$ (firm
lines) and the zero flux limit of the SD5-brane with $d_0=3,
d_1=6$ and $p^1=0.5$ (dash lines).}
\end{figure*}

\subsubsection{\label{sec:4.1.2}\bf{Ricci-flat external space (SM2 and SD2 branes)}}
It can be easily seen from Eq. \rf{3.11} that the external space
$M_1$ expends only in the case $\xi_1>0\rightarrow p^1>0$. Because
$p^1\leq 1 \Rightarrow [(D-2)/d_0d_1](\xi_1/\xi_2) =
[(D-2)/d_0d_1]^{1/2}p^1<1$, then the deceleration parameter of the
external space $q_1 >0$ for all times $\tau \in (-\infty ,0]$ (see
Eq. \rf{3.12}) and the acceleration is absent. Additionally, the
internal space $M_0$ expands to infinity at late times  which
obviously contradicts the observations. Thus, this case is not of
interest for us.

\subsection{\label{sec:4.2}Einstein frame}
Now, we investigate the dynamical behavior of the corresponding
S$p$-branes in the Einstein frame splitting our consideration into
two separate cases depending on the choice of the external factor
space.

\subsubsection{\label{sec:4.2.1}\bf{Hyperbolic external space (SM6 and SD5 branes)}}
In this case we obtain the following expressions:
\be{3.15}
\Omega  = a_1^{ - \frac{d_1 }{d_0-1}} = \exp ( - \frac{\xi _1
}{d_0-1}\tau )
\ee
for the conformal factor,
\be{3.16}
f(\tau ) = f_{-E(0)} (\tau ) = \Omega ^{ - 1} e^{\gamma _0 } =
\left[ {\sinh(  \xi _2 |\tau |)} \right]^{-\frac{d_0}{d_0-1}}
\ee
for the function $f(\tau )$ and
\be{3.17}
\tilde {a}_0 (\tau ) = \Omega^{-1}a_0 = g_{-}^{-1} = \left[
{\sinh( \xi _2 |\tau |)} \right]^{-\frac{1}{d_0-1}}
\ee
for the scale factor of the external space. Here, we consider the
interval $(-\infty ,0]$ of the harmonic time $\tau $ which is
mapped onto the interval $[0,+\infty )$ of the synchronous time
$\tilde t$. Thus, the Hubble and the deceleration parameters of
the external factor space $M_0$ read:
\ba{3.18}
\tilde {H}_0 &=&  \frac{\xi_2}{(d_0-1)f_{-E(0)} }\coth\left( {\xi
_2 |\tau |}
\right)\, , \\
\label{3.19}   \tilde {q}_0 &=&  \frac{\xi _2^2
}{(d_0-1)f_{-E(0)}^2 (\tau )}\, .
\ea
These equations clearly show that the
expanding external space is decelerating one because $\tilde {H}_0
>0\, ,  \tilde {q}_0 >0 \, \, \forall \, \, \tau \in (-\infty ,0]$.

\subsubsection{\label{sec:4.2.2}\bf{Ricci-flat external space (SM2 and SD2 branes)}}
Let now the factor space $M_1$ be the external space. For this
choice of the external space the conformal factor reads
\ba{3.20} \nonumber\Omega (\tau ) &= & a_0^{-\frac{d_0}{d_1-1}}=
\exp
\biglb( {\frac{d_0 }{(d_0 - 1)(d_1-1)}\xi _1 \tau } \bigrb)\\
&\times&\biglb[ {\sinh( \xi _2 |\tau |)} \bigrb]^{\frac{d_0 }{(d_0
- 1)(d_1-1)}}\, . \ea
With the help of this expression we can define the function
$f(\tau )$
\ba{3.21}  &f&(\tau ) = f_{-E(1)} (\tau ) = \Omega _1^{ - 1}
e^{\gamma _0 }\\\nonumber & = & \exp \biglb( {  \frac{1-d_0
-d_1}{(d_0 - 1)(d_1-1)}\xi _1 \tau } \bigrb) \biglb[ {\sinh\biglb(
{ \xi _2 |\tau |} \bigrb)} \bigrb]^{-\frac{d_0d_1 }{(d_0 -
1)(d_1-1)}} \ea
and the scale factor $\tilde a_1(\tau )$
\ba{3.22}\nonumber \tilde {a}_1 (\tau ) &=& \Omega^{-1}a_1 = \exp
\biglb( { \frac{1-d_0 -d_1}{d_1(d_0 - 1)(d_1-1)}\xi _1 \tau }
\bigrb)\\ &\times& \biglb[ {\sinh\biglb( { \xi _2 |\tau |}
\bigrb)} \bigrb]^{-\frac{d_0 }{(d_0 - 1)(d_1-1)}}\, . \ea
As for the internal space scale factor $a_0(\tau )$, it has the
form \rf{1.8} with the asymptotes \rf{3.4} and \rf{3.5}.

Similar to the previous case, we choose the interval $(-\infty
,0]$ of the harmonic time $\tau $. It can be easily verified that
this interval is mapped onto the interval $[0,+\infty )$ of the
synchronous time $\tilde t$. It is of interest to get the late
time asymptotes for the scale factors. To get them, we obtain
first the relation between the synchronous and harmonic times at
late stages:
\ba{3.23}\nonumber \xi_2 \tilde t &=&
\frac{(d_0-1)(d_1-1)}{d_0+d_1-1}\biglb(\xi_2|\tau
|\bigrb)^{-\frac{d_0+d_1-1}{(d_0-1)(d_1-1)}} \, , \\
 &\tau& \rightarrow -0 \Rightarrow t \rightarrow +\infty\, , \ea
which enable us to write the late time asymptotes in both gauges:
\ba{3.24} \tilde a_1 &\simeq& \biglb(\xi_2|\tau
|\bigrb)^{-\frac{d_0}{(d_0-1)(d_1-1)}}\, \\\nonumber &\simeq& \,
\biglb[\frac{d_0+d_1-1}{(d_0-1)(d_1-1)}\, \xi_2\tilde t\,
\bigrb]^{\frac{d_0}{d_0+d_1-1}}\, ,\\
\label{3.25} a_0 &\simeq& \biglb(\xi_2|\tau
|\bigrb)^{-\frac{1}{d_0-1}}\, \\\nonumber &\simeq& \,
\biglb[\frac{d_0+d_1-1}{(d_0-1)(d_1-1)}\, \xi_2\tilde t\,
\bigrb]^{\frac{d_1-1}{d_0+d_1-1}}\, . \ea
Thus, both the external and the internal scale factors expand at
late times. However, the rate of the expansion of the internal
space $M_0$ is less than for the external space $M_1$. For
example, in the case $d_1=3,\, d_0=6$ we get: $\tilde a_1 \sim
\tilde t^{\, 3/4}$ and $a_0 \sim \tilde t^{\, 1/4}$. So, in spite
of this expansion, we suppose that the internal scale factor is
still less than the Fermi length which makes it unobservable at
present time.

To investigate the accelerating behavior of the external space
$M_1$, let us define its Hubble and deceleration parameters:
\begin{widetext}
\be{3.26} \tilde {H}_1 (\tau ) = - \frac{\biglb( {(D - 2)\xi _1 +
d_0d_1 \xi _2 \coth(\xi _2 \tau )} \bigrb)}{d_1(d_1-1)(d_0 -
1)f_{-E(1)} (\tau )} = - \frac{1}{(d_1-1) f_{-E(1)} (\tau )}\sqrt
{\frac{D-2 }{d_1(d_0-1) }} \, \, m(\tau )\, \ee
and
\ba{3.27} \tilde {q}_1 (\tau ) &=& \frac{ {\biglb[ {(D - 2)\xi _1
+ d_0d_1 \xi _2 \coth(\xi _2 \tau )} \bigrb]^2 - d_0(d_0 -
1)d_1^2\xi_2^2\sinh^{-2}(\xi _2 \tau )} }{d_1^2(d_1-1)(d_0 -
1)^2f_{-E(1)}^2 (\tau )} \\&=& \frac{  (D - 2) m^2(\tau ) - {(d_0
- 1)d_1\sinh^{-2}(\xi _2 \tau )} }{d_1(d_1-1)(d_0 - 1)f_{-E(1)}^2
(\tau )} \, , \nonumber \ea
\end{widetext}
where (see also Refs. \cite{TW,Roy})
\be{3.28} m(\tau ) := {p^1 +
\sqrt {\frac{d_0d_1 }{D - 2}} \coth(\xi _2 \tau )}\, .
\ee
It can be easily seen that this function is negative: $m(\tau )< 0
\, \, \forall \, \, \tau \in (-\infty ,0]$ because $|p^1|\leq 1$.
Thus, starting from zero value\footnote{For $\tau \rightarrow
-\infty $ irrespective of the sign of $\xi_1$ eq. \rf{3.22} has
the asymptote $\tilde a_1 \sim \exp
\biglb\{\frac{d_0}{(d_0-1)(d_1-1)}\xi_2|\tau
|\biglb(\frac{D-2}{d_1d_0}\frac{\xi_1}{\xi_2}-1\bigrb)\bigrb\}
\rightarrow 0 $ .} the external space $M_1$ expands for all times
(see Eq. \rf{3.26}). From other side, the condition of its
acceleration reads
\ba{3.29}\nonumber \frac{d_1}{D - 2}\coth^2(\xi _2 \tau )&+&
2\sqrt {\frac{d_0d_1 }{D - 2}}p^1\coth(\xi _2 \tau )\\ + (p^1)^2
&+& {\frac{(d_0-1)d_1 }{D - 2}} < 0\, . \ea
Because $\coth(\xi _2 \tau )<0$ for $\tau \in (-\infty ,0]$, this
inequality is possible only for positive values of the parameter
$p^1:\, p^1>0$. Moreover, the corresponding quadratic equation
should have two roots defining the harmonic time of the beginning
($\tau_{(a)start}$) and ending ($\tau_{(a)fin}$) of the
acceleration. For these roots we obtain the following relation:
\ba{3.30} &\coth&(\xi_2\tau_{(a)start}) -
\coth(\xi_2\tau_{(a)fin})
\\ \nonumber &=&
2\sqrt{\frac{(d_0-1)(D-2)}{d_1}}\sqrt{(p^1)^2-\frac{d_1}{D-2}}\, .
\ea
This difference is positive because $\coth$ is a monotone
decreasing function. So the stage of the acceleration takes place
only if the parameter $p^1$ satisfies the inequality
\be{3.31} (p^1)^2>\frac{d_1}{D-2} \, . \ee


For $p^1=1$ we restore the results of the paper \cite{TW}.
However, a new result is that scalar field with $(d_0-1)/(D-2)
\leq (p^2)^2\leq 1$ prevents the acceleration. In Fig. 4, we
present different behavior of the external $\tilde a_1$ and
internal $a_0$ scale factors as well as the deceleration parameter
$-\tilde q_1$ of the external space $M_1$ depending on the choice
of the parameter $p^1$. The firm lines corresponds to the values
of $p^1$ satisfying the condition of the acceleration \rf{3.31}.
$t^{acc}_{in}\, \, \mbox{and}\, \, t^{acc}_{fin}$ denote
respectively the times of the beginning and ending of the external
space acceleration. The dash lines correspond to the case when the
parameter $p^1$ does not satisfy the condition \rf{3.31} and the
stage of the acceleration is absent.

\begin{figure*}[htbp]

    \includegraphics[width=6.5in,height=3.6in]{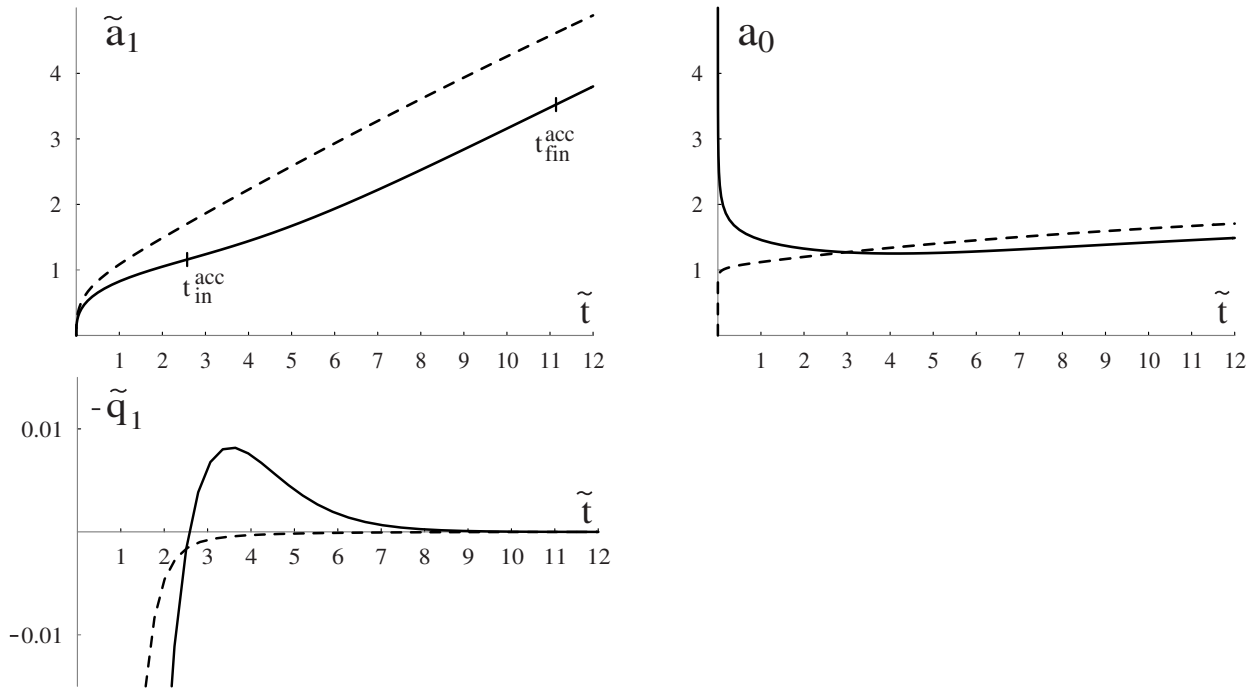}

\caption{Typical form of the external $\tilde a_1$ and internal
$a_0$ scale factors as well as the deceleration parameter $-\tilde
q_1$ of the external Ricci-flat factor space $M_1$ in Einstein
frame (synchronous time gauge). Specifically, the firm lines
represent the vacuum limit of the accelerating SM2-brane with
$d_0=7, d_1=3$ and $p^1=1$, and the dash lines correspond to the
zero flux limit of the decelerating SD2-brane with $d_0=6, d_1=3$
and $p^1=0.5$.}
\end{figure*}



\section{\label{sec:5}Variation of the fine structure constant }
\setcounter{equation}{0}

Above, we considered the model with the dynamical internal spaces.
It is well known that the internal space dynamics results in the
variation of the fundamental constants such as the gravitational
constant and the fine structure constant (see, e.g., Refs.
\cite{Uzan,Kub}). For example, the effective four-dimensional
fine-structure constant
is inversely proportional to the volume of the internal space
(see, e.g., Refs. \cite{GSZ,CV,Kub}): $\alpha  \sim V^{-1}_{(I)}
\sim a_{(I)}^{-d_{(I)}}$. Here, the indices "I" and "E" denote the
internal and external spaces correspondingly. The origin of such
dependence can be easily seen if we add a higher-dimensional
electromagnetic action (which should not affect the investigated
above dynamics of the model) and perform the dimensional reduction
to an effective four-dimensional theory. It results in the term of
the form $\sqrt{g^{(E)}}\, (V_{(I)}/e^2) F^2$ \cite{GSZ,Kub} which
leads to the indicated above dependence for the effective
fine-structure constant. Thus, if $V_{(I)}$ is a dynamical
function which varies with time then the effective
four-dimensional constants will vary as well. For the
fine-structure constant, such variations take place in both frames
because the quantity $\sqrt{g^{(E)}}\, (V_{(I)}/e^2) F^2$ is
invariant in four-dimensional space-time with respect to the
conformal transformation of the metric $g^{(E)}$. Therefore, in
both frames we arrive to the following expression for the
variation of $\alpha$:
\be{4.1}
\left|\frac{\dot{\alpha}}{\alpha}\right|=
\left|\frac{\dot{V}_{(I)}}{V_{(I)}}\right|=
\left|d_{(I)}H_{(I)}\right|\, ,
\ee
where the dot denotes the synchronous time derivatives and
$H_{(I)} = \dot a_{(I)}/a_{(I)}$.

There are strong constraints on $\dot{\alpha}/\alpha$ from a
number of experimental and observational considerations
\cite{Uzan}. For our calculations we use the estimate
$|\dot{\alpha}/\alpha| \lesssim 10^{-15}\mbox{yr}^{-1}$
\cite{Webb} which follows from observations of the spectra of
quasars. Combining this with the accepted value for the current
Hubble rate $H_{(E)} = \dot a_{(E)}/a_{(E)} \sim
10^{-10}\mbox{yr}^{-1}$ leads to
\be{4.2}
\left|\frac{H_{(I)}}{H_{(E)}}\right|\lesssim 10^{-5}\, .
\ee

Let us test now the models from sections \ref{sec:3} and
\ref{sec:4} for the purpose of their satisfaction of the condition
\rf{4.2}. We perform this investigation only for the cases with
the acceleration of the external space.

\subsection{\label{sec:5.1}Brans-Dicke frame}

\subsubsection{\label{sec:5.1.1}Spherical space}

In this case the Hubble parameters for the factor spaces are given
by Eqs. \rf{2.9} and \rf{2.11}. Therefore, depending on the choice
of the external space, we obtain the following results:

\vspace{0.5cm}

1.{\it{ spherical external space (section \ref {sec:3.1.1})}}

\be{4.3}
\left|\frac{H_{(I)}}{H_{(E)}}\right| =
\left|\frac{H_1}{H_0}\right| = \frac{d_0-1}{d_1\left|{1+
\frac{\xi_2}{\xi_1}\tanh(\xi_2\tau)}\right| }\sim \mathcal{O}(1)\,
.
\ee
This estimate arises from the condition \rf{2.14}. Therefore, in
this case we arrive to the obvious contradiction with the
experimental bounds.

\vspace{0.5cm}

2. {\it{Ricci-flat external space (section \ref {sec:3.1.2})}}

\be{4.4}
\left|\frac{H_{(I)}}{H_{(E)}}\right| =
\left|\frac{H_0}{H_1}\right| = \frac{d_1\left|{1+
\frac{\xi_2}{\xi_1}\tanh(\xi_2\tau)}\right| }{d_0-1}\sim
\mathcal{O}(1)\, .
\ee

This estimate is valid for all times $\tau \in (-\infty ,+\infty )
$. For $\xi_1/\xi_2 <1$, the only exclusion is a very short period
of time in the vicinity of the turning point $\tanh (\xi_2\tau) =
-\xi_1/\xi_2$ of the internal space $M_0$, i.e. for the times
$\tanh (\xi_2\tau) \in [-\xi_1/\xi_2 - \delta ,-\xi_1/\xi_2 +
\delta ]$ with $\delta \sim (\xi_1/\xi_2 )10^{-5}$.
In general, this model conflicts with the observations.


\subsubsection{\label{sec:5.1.2} Hyperbolic space}

In this case the Hubble parameters for the factor spaces are given
by Eqs. \rf{3.9} and \rf{3.11}. Here, the acceleration takes place
only in the case of the hyperbolic external space.

\vspace{0.5cm}

1.{\it{ hyperbolic external space (section \ref {sec:4.1.1})}}

\vspace{0.5cm}

With the help of Eqs. \rf{3.9} and \rf{3.11}, the ratio between
the Hubble parameters is given by
\be{4.5}
\left|\frac{H_{(I)}}{H_{(E)}}\right| =
\left|\frac{H_1}{H_0}\right| = \frac{d_0-1}{d_1\left|{1+
\frac{\xi_2}{\xi_1}\coth(\xi_2\tau)}\right| }\, .
\ee

As we have seen in section \ref{sec:4.1.1}, there are two
distinguishing scenarios in this case. The first scenario
corresponds to $\xi_1>\xi_2$ (it happens, e.g. in the case of the
absence of scalar field: $p^1 =1$). As for this particular case
$\xi_2/\xi_1 \sim \mathcal{O}(1)$, we can achieve the necessary
smallness of the ratio \rf{4.2} for late times:
\be{4.6}
\left|\frac{H_1}{H_0}\right| < 10^{-5} \quad \mbox{for} \quad
|\xi_2\tau | < 10^{-5} \, , \quad (\xi_1>\xi_2)\, .
\ee
The second scenario takes place if $0<\xi_1\leq \xi_2$. It can be
easily seen that the condition \rf{4.2} is satisfied for small
parameter $\xi_1$:
\be{4.7}
\left|\frac{H_1}{H_0}\right| < 10^{-5} \quad \forall  \quad \tau
\, \in \,  (-\infty ,0]  \, , \quad (\xi_1/\xi_2 \lesssim
10^{-5})\, .
\ee
We can weaken the condition $\xi_1/\xi_2 \lesssim 10^{-5}$ if
demand the execution of the condition \rf{4.2} from the time
$\tau_a$ of the beginning of the acceleration (see Eq. \rf{3.14}):
\be{4.8}
\left|\frac{H_1}{H_0}\right| < 10^{-5} \quad \mbox{for}  \quad
\tau_a \leqslant \tau \leqslant 0\,  , \quad (\xi_1/\xi_2 \lesssim
10^{-5/2})\, .
\ee
Therefore, in the case of the hyperbolic external case we can
satisfy the condition \rf{4.2} either for sufficiently late times
$|\xi_2\tau | < 10^{-5} $ or for very dynamical scalar field which
results in the smallness of the parameter $p^1$:
$\sqrt{d_0d_1/(D-2)}p^1 \lesssim 10^{-5}$ for \rf{4.7} or
$\sqrt{d_0d_1/(D-2)}p^1 \lesssim 10^{-5/2}$ for \rf{4.8}.

\subsection{\label{sec:5.2}Einstein frame}

In the Einstein frame, there is only one case with the
accelerating stage for the external space. It describes the model
with Ricci-flat external and hyperbolic internal spaces.

\subsubsection{\label{sec:5.2.1}Hyperbolic internal space (section \ref{sec:4.2.2})}


In this case the Hubble parameter of the external space $M_1$ is
defined by Eq. \rf{3.26}. Concerning the Hubble parameter of the
internal factor space $M_0$, it is necessary to perform the
evident substitution $f_{-BD}(\tau) \rightarrow f_{-E(1)}(\tau)$
in formula \rf{3.9} because in the Einstein frame the function
$f(\tau)$ in Eq. \rf{1.14} is defined by $f_{-E(1)}(\tau)$. Thus,
the ratio of the Hubble parameters reads
\ba{4.9} \left|\frac{H_{(I)}}{H_{(E)}}\right| &=&
\left|\frac{H_0}{\tilde {H}_1}\right| \\ \nonumber &=&
\frac{d_1(d_1-1)}{D-2}
\left|\frac{\frac{\xi_2}{\xi_1}+\tanh(\xi_2\tau)}{\frac{d_0d_1}{D-2}
\frac{\xi_2}{\xi_1}+\tanh(\xi_2\tau)}\right| \sim \mathcal{O}(1)\,
. \ea
This estimate is valid for all times $\tau \in (-\infty ,0]$. For
$\xi_2/\xi_1 <1$, the only exclusion is a very short period of
time in the vicinity of the turning point $\tanh (\xi_2\tau) =
-\xi_2/\xi_1$ of the internal space $M_0$, i.e. for the times
$\tanh (\xi_2\tau) \in [-\xi_2/\xi_1 - \delta ,-\xi_2/\xi_1 +
\delta ]$ with $\delta \sim 10^{-5}$.
Therefore, in general, this model conflicts with the experimental
bounds.

\subsection{\label{sec:5.3}Static internal space}

It is clear that the effective fundamental constants do not
variate if the internal space is static ("frozen"). Additionally,
it results in the equivalence between Brans-Dicke and Einstein
frames. In our model it takes place only if the parameter $\xi_1
=0 \Rightarrow p^1 = 0$ (see, e.g. Eqs. \rf{1.8}-\rf{1.10}), i.e.
when the factor space $M_1$ plays the role of the internal space.
Let us investigate this possibility in more details.

First, we consider the {\it{spherical}} external space. It follows
from Eq. \rf{2.10} that the external space $M_0$ is decelerating
because $q_0>0$ for $\xi_1=0$. Moreover, the static solution is
unstable. To see it, let us suppose that the internal space scale
factor $a_1$ be freezed up to an arbitrary time $\tau_0$. Then,
small fluctuations $\delta \xi_1 =
\left[d_1(d_0-1)/(D-2)\right]^{1/2}\delta p^1$ results in the
following dynamics:
\be{4.10} \left.a_1(\tau )\right|_{\tau \ge \tau_0} = \exp
[(\delta\xi_1 /d_1)(\tau -\tau_0)] \, ,\quad \tau \in
[\tau_0,+\infty ) \ee
(see Eq. \rf{1.8}). Thus, the scale factor $a_1$ goes from the
constant value either to $+\infty$ (for positive $\delta p^1$) or
to zero (for negative $\delta p^1$). At the same time, the
external scale factor $a_0$ remains decelerating because the small
fluctuation $\delta \xi_1$ does not satisfy the acceleration
condition \rf{2.14}. Therefore, this case is not of interest for
us.

Second, we turn to the {\it{hyperbolic}} external space. Here, the
external space $M_0$ is again decelerating (see Eq. \rf{3.10} for
$\xi_1=0$). Because of small fluctuations $\delta \xi_1$ at an
arbitrary moment $\tau_0<0$, the scale factor $a_1$ acquires the
dynamics:
\be{4.11} \left.a_1(\tau )\right|_{\tau \ge \tau_0} = \exp
[(\delta\xi_1 /d_1)(\tau -\tau_0)] \, ,\quad \tau \in [\tau_0,0]\,
. \ee
Thus, for negative $\delta \xi_1$ the internal scale factor
approaches asymptotically the value $\exp
[-(|\delta\xi_1|/d_1)|\tau_0|]$ and the external space remains
decelerating. In this case the internal space varies in finite
limits of the order of $\mathcal{O}(1)$ (from this point we can
call this case "quasi stable"). For positive $\delta \xi_1$, the
internal scale factor approaches asymptotically the value $\exp
[(\delta\xi_1/d_1)|\tau_0|]$ and the external space starts to
accelerate at the time $\coth(\xi_2 \tau_a) = -\xi_2/\delta\xi_1$
(see Eq. \rf{3.14}). The case of the positive $\delta \xi_1$ is of
interest because, first, the external space begins to accelerate,
and, second, the variations of the fundamental constant do not
contradict the observations if the ratio $\delta \xi_1/\xi_2$
satisfies the conditions similar to those for the ratio
$\xi_1/\xi_2$ in the expressions \rf{4.7} and \rf{4.8}. However,
the scale factor $a_1$ can considerably increase if
$(\delta\xi_1/d_1)|\tau_0|>>1$. In this case the solution is
unstable.



\section{\label{sec:6}Fixation of the fine structure constant }
\setcounter{equation}{0}

Let us consider now the case of three factor spaces with the
topology of the manifold of the form: $M=\mathbb{R}\times M_0
\times \mathbb{R}^{d_1} \times \mathbb{R}^{d_2}$ where $M_0$ is
$d_0$-dimensional spherical ($S^{d_0}$) or hyperbolic ($H^{d_0}$)
space.

Here, the solution (in the Brans-Dicke frame) is (see Refs.
\cite{BZpositive,BZnegative})
\ba{5.1} a_0(\tau)&=&A_0
\exp\biglb(-\frac{\xi_1}{d_0-1}\tau\bigrb)
\times \frac{1}{\tilde g_{\pm}(\tau )}\, , \\
\label{5.2} a_{1}(\tau)&=&A_1
\exp\Biglb({\biglb(\frac{\xi_1}{d_1+d_2}-
\frac{\xi_3}{d_1}\bigrb)\tau}\Bigrb) \, , \\
\label{5.3} a_{2}(\tau)&=&A_2
\exp\Biglb({\biglb(\frac{\xi_1}{d_1+d_2}+
\frac{\xi_3}{d_2}\bigrb)\tau}\Bigrb) \, ,\\
\label{5.4} \varphi(\tau)&=&p^3\tau+q \, ,
 \ea
where
\ba{5.5} \tilde g_+ (\tau )&=&
\left(\frac{2\varepsilon}{R_0}\right)^{-1/2(d_0-1)}
\cosh^{1/(d_0-1)}(\xi_2\tau)\, \nonumber, \\ \quad &-\infty & <
\tau <+\infty \, ,
 \quad \varepsilon >
 0\,
 \ea
and
\ba{5.6} \tilde g_ - (\tau
)&=&\left(\frac{2\varepsilon}{|R_0|}\right)^{-1/2(d_0-1)}
\sinh^{1/(d_0 - 1)}\left( {\xi _2 |\tau |} \right)\, , \nonumber
\\\quad &|\tau |&>0\, ,
 \quad \varepsilon \ge
 0\, .
\ea
As usual in this paper, the index $+ (-)$ indicates that
considered formula is related to the spherical (hyperbolic) factor
space $M_0$. In the case $\varepsilon =0$, Eq. \rf{5.6} is reduced
to the form
\be{5.7}
\tilde g_-(\tau )
= \biglb[(d_0-1)|\tau |\bigrb]^{1/(d_0-1)}\, , \ee
where we used the formula $|R_0|=d_0(d_0-1)$.

In Eqs. \rf{5.1}-\rf{5.6},
\ba{5.8}
\xi_1&=&\sqrt{\frac{(d_1+d_2)(d_0-1)}{(D-2)}}\, \, p^1\, ,\nn \\
\xi_2&=&\sqrt{\frac{d_0-1}{d_0}\, \, 2\varepsilon}\, , \\
\xi_3&=& \sqrt{\frac{d_1d_2}{d_1+d_2}}\, \, p^2\, \nn
\ea
and
\be{5.9}
2\varepsilon=(p^1)^2+(p^2)^2+(p^3)^2\, .
\ee
Parameters, $A_0, A_1, A_2, p^1, p^2, p^3$ and $q$ are the
constants of integration with the following constraint:
$A_0^{d_0}A_1^{d_1}A_2^{d_2}=A_0$.

For this solution, the Hubble and deceleration parameters read:
\ba {5.10} H_{\pm 0}(\tau) & = &  -
\frac{1}{f_{\pm}(\tau)}\frac{\xi_1+\xi_2h_{\pm}(\tau)}{d_0-1}\, , \\
\label{5.11} q_{\pm 0}(\tau ) & = & \frac{\xi_2}{f_{\pm
}^2(\tau)}\frac{\xi_2+\xi_1h_{\pm}(\tau)}{d_0-1}\, ,\\
\label{5.12} H_{\pm
1}(\tau)&=&\frac{1}{f_{\pm}(\tau)}\Biglb(\frac{\xi_1}{d_1+d_2}
-\frac{\xi_3}{d_1}\Bigrb),
\\\label{5.13}
H_{\pm
2}(\tau)&=&\frac{1}{f_{\pm}(\tau)}\Biglb(\frac{\xi_1}{d_1+d_2}+\frac{\xi_3}{d_2}\Bigrb)\,
, \ea
where the transition function $f(\tau )$ (see Eq. \rf{1.13}) is
\ba{5.14} f_\pm(\tau)&=&e^{\gamma_0}=a_0^{d_0}a_1^{d_1}a_2^{d_2}
\\ &=& A_0\exp(-\frac{\xi_1}{d_0-1}\, \tau)\times \frac{1}{\tilde
g_\pm(\tau)^{d_0}}\,\nonumber \ea
and
\be{5.15}
h_{\pm} (\tau )= \left\{
\begin{array}{c}
  \tanh (\xi_2\tau )\, , \\
  \vphantom{\exp } \\
  \coth (\xi_2\tau )\, ,
\end{array}\right. \quad \begin{array}{c}
  \tau \in (-\infty ,+\infty )\, , \quad  R_0>0 \, , \\
  \vphantom{\exp } \\
  \tau \in (-\infty ,0]\, , \quad  R_0<0 \, .
\end{array}
\ee

In this section the factor space $M_0$ is treated as the external
one. This choice is justified below. As it follows from Eqs.
\rf{5.10}-\rf{5.15}, the dynamics of the model is similar to that
described in sections \ref{sec:3.1.1} and \ref{sec:4.1.1}. For
example, the {\it{spherical}} external space $M_0$ undergoes the
accelerating expansion (during the period \rf{2.13}) and both
internal spaces $M_1$ and $M_2$ contract if $\xi_1<0$ and
$\xi_3<(d_2/(d_1+d_2))|\xi_1|$ for positive $\xi_3>0$ or
$|\xi_3|<(d_1/(d_1+d_2))|\xi_1|$ for negative $\xi_3<0$. In the
case of the {\it{hyperbolic}} external space, the accelerating
expansion of $M_0$ is possible only if $\xi_1$ is positive:
$\xi_1>0$. Here, the acceleration of $M_0$ is either eternal (if
$\xi_1>\xi_2$) or starts at the time \rf{3.14} (if $0<\xi_1\leq
\xi_2$). Concerning the internal spaces $M_1$ and $M_2$ we can say
that at least one of them expands approaching the finite value
$A_1$ or $A_2$.

As to the variations of the effective fine structure constant, we
obtain
\be{5.16}
\left|\frac{\dot{\alpha}}{\alpha}\right|=
\left|\frac{\dot{V}_{(I)}}{V_{(I)}}\right|=
\left|d_{1}H_{1}+d_{2}H_{2}\right|\, ,
\ee
where $V_{(I)} \sim a_1^{d_1}a_2^{d_2}$. Since the combination
$d_1H_1$ in the case of one internal space gives exactly the same
expression as the combination $d_1H_1+d_2H_2$ in the case of two
internal spaces (see Refs. \rf{2.11}, \rf{3.11}, \rf{5.12} and
\rf{5.13}), we arrive to the conclusions with respect to the
variations of $\alpha$ similar to those obtained in sections
\ref{sec:5.1.1} and \ref{sec:5.1.2} : the spherical model is in
conflict with the observations (see Eq. \rf{4.3}) and the
hyperbolic model can be in agreement with the experimental bounds
either at very late times (see Eq. \rf{4.6}) or for very small
$\xi_1$ (see Eqs. \rf{4.7} and \rf{4.8}).

Obviously, the effective four-dimensional fundamental constants
are fixed if the total volume of the internal spaces is constant.
Now, we try to answer the following question. Is it possible to
construct the model with dynamical scale factors but fixed total
volume of the internal spaces? The simple analysis of Eqs.
\rf{5.1}-\rf{5.3} shows that such possibility exists only if we
chose the Ricci-flat factor spaces $\mathbb{R}^{d_1}$ and
$\mathbb{R}^{d_2}$ as the internal ones and put $p^1=0$. In this
case
\be{5.17}
V_{(I)} \sim \left. a_1^{d_1}a_2^{d_2}\right|_{p^1=0} =
A_1^{d_1}A_2^{d_2} = \mbox{const}\, .
\ee
Hence, in spite of the dynamical behavior of the internal scale
factors, first, the Brans-Dicke and Einstein frames are equivalent
each other and, second, the fundamental constants are fixed. It
was the main reason to chose the factor space $M_0$ as the
external one. At first sight, this model looks very promising.
However, it has a number of drawbacks. First, the external space
$M_0$ is the decelerating one: $q_{\pm 0}(\tau )>0$ (see Eq.
\rf{5.11}).

Additionally, it is necessary to investigate this model for the
purpose of its stability with respect to the fluctuations of the
parameter $p^1$. It can be easily seen that due to small
fluctuations $\delta\xi_1 = [(d_1+d_2)(d_0-1)/(D-2)]^{1/2}\delta
p^1$ at an arbitrary moment $\tau_0$ the internal volume acquires
the following dynamics:
\be {5.18}
V_{(I)}=A_1^{d_1}A_2^{d_2}\exp(\delta\xi_1(\tau-\tau_0))\, ,
\ee
where $\tau\in [\tau_0, +\infty)$ for the spherical $M_0$ and
$\tau\in [\tau_0, 0]$ for the hyperbolic $M_0$. Thus, the
stability analysis can be performed in full analogy with section
\ref{sec:5.3} of the static internal space. We obtain that the
case of spherical external space is unstable with the decelerating
behavior and the case of the hyperbolic external space is "quasi
stable" for $\delta\xi_1<0$ and unstable for $\delta\xi_1>0$. In
the later case the factor space $M_0$ can acquire the stage of the
acceleration without too high variation of $\alpha$.

To conclude this section, we consider a particular model with
fixed internal volume \rf{5.17} and additional condition
$\varepsilon =0$. It takes place if scalar field is an imaginary,
i.e. $\varphi$ is a phantom field (see, e.g., \cite{Gibb,SSD,DSS}
and numerous references therein). For the
hyperbolic\footnote{$\vphantom{\exp}$ Classical Lorentzian
solutions with $\varepsilon =0$ exist only for the hyperbolic
$M_0$.} external space $M_0$ the solution (in the harmonic time
gauge) is given by Eqs. \rf{5.1}-\rf{5.4} with the following
substitution: $\xi_1=0,\, p^3=ip^2$ and $\tilde g_{-}$ from Eq.
\rf{5.7}.  This particular model is of interest because of its
integrability in the synchronous time gauge where the solution
reads
 \ba{5.19} a_0(t)&=& t \, ,\\
\label{5.20} a_{1}(t)&=&A_1 \exp\biglb(\frac{\xi_3}{d_1(d_0-1)}
\biglb(\frac{A_0}{t}\bigrb)^{d_0-1}\bigrb)\, ,
\\ \label{5.21}
a_{2}(t)&=&A_2 \exp\biglb(-\frac{\xi_3}{d_2(d_0-1)}
\biglb(\frac{A_0}{t}\bigrb)^{d_0-1}
\bigrb)\, ,\\
\label{5.22} \varphi (\tau ) &=&
i\frac{p^2}{d_0-1}\biglb(\frac{A_0}{t}\bigrb)^{d_0-1}+q\, \ea
and $t\in [0,+\infty )$. Hence, the scale factor of the external
space behaves as in the case of the Milne solution with zero
acceleration. This is a transitional case between the accelerating
and decelerating behavior. Any perturbations $\delta p^1$ result
in non-zero $2\varepsilon=(\delta p^1)^2>0$. The behavior of such
perturbed model is described by Eqs. \rf{5.1}-\rf{5.3} with $\xi_1
\rightarrow \delta\xi_1 = [(d_1+d_2)(d_0-1)/(D-2)]^{1/2}\delta
p^1$ and $2\varepsilon=(\delta p^1)^2$. In this case
$|\delta\xi_1/\xi_2|=\sqrt{d_0(d_1+d_2)/(D-2)}>1$. Thus, for
positive fluctuations $\delta \xi_1$ the external space $M_0$
undergoes the eternal acceleration in accordance with the results
of section \ref{sec:4.1.1}. However, the variations of $\alpha$ do
not contradict the experimental bounds only for very late times,
as we have seen in section \ref{sec:5.1.2}. Additionally, the
internal space volume $V_{(I)}$ can considerably increase if
$\delta\xi_1|\tau_0|>>1$ (see Eq. \rf{5.18}). Therefore, this
solution is unstable.


\section{\label{sec:7}Conclusion}

In the present paper we investigated the possibility of generating
the late time acceleration of the Universe from gravity on product
spaces with only one non-Ricci-flat factor space. The model
contains minimally coupled free scalar field as a matter source.
Dynamical solutions for this model are called S-brane (spacelike
brane) solutions. The analysis was performed in the Brans-Dicke
and Einstein frames.
We found that in the context of considered models, non-Einsteinian
gravity provides more possibilities for accelerating cosmologies
than the Einsteinian one. As we already mentioned in the
Introduction, such different behavior of the external space scale
factors in both of these frames is not surprising because these
scale factors are described by different variables connected with
each other via the conformal transformation (see, e.g., Eq.
\rf{a.5} in Appendix). Moreover, the synchronous times in both of
these frames are also different. As a consequence of these
discrepancies, the scale factors of the external space in both
frames behave differently. In the Brans-Dicke frame, stages of the
accelerating expansion exist for all types of the external space
(flat, spherical and hyperbolic). However, in the Einstein frame,
the model with flat external space and hyperbolic compactification
of the internal space is the only one with the stage of the
accelerating expansion. The reason for this acceleration is rather
clear. After dimensional reduction of the considered models and
conformal transformation to the Einstein frame, we obtain an
effective potential of the form: $U = - (1/2)e^{2\gamma _0 }R_0
e^{ - 2\beta ^0}$ (see Eq. \rf{1.7}), which plays the role of an
effective cosmological "constant". Thus, the acceleration is
possible only if the internal space curvature $R_0<0$. The
presence of minimally coupled free scalar field does not help the
acceleration because this field does not contribute into the
potential. Nevertheless, it make sense to include such field into
the model because it results in more reach and interesting
dynamical behavior\footnote{We have seen that dynamical picture of
the model considerably depends on the relation between parameters
$\xi_1$ and $\xi_2$ introduced in Eqs. \rf{1.8}-\rf{1.10}. If
scalar field is absent, $|\xi_1|/\xi_2 =
\left[d_1d_0/(D-2)\right]^{1/2}>1$. However, in the presence of
scalar field this ratio is not fixed but varies in the limits $0
\leq |\xi_1|/\xi_2 \leq \left[d_1d_0/(D-2)\right]^{1/2}$.}.
Moreover, we have seen in section \ref{sec:4.2.2} that scalar
field
can prevent the
acceleration in the Einstein frame. This is a new result in
comparison with Refs. \cite{TW,CHNOW}.

It is well known that the dynamical behavior of the internal
spaces results in the variation of the effective four-dimensional
fundamental constants. Therefore, we investigated the rate of
variation of the fine structure constant for the cases of the
accelerating external spaces. It was shown that the case of the
hyperbolic external space in the Brans-Dicke frame is the only
model which can satisfy the experimental bounds for the fine
structure constant variations.

It is clear that the fundamental constant variations are absent if
the total volume of the internal spaces is constant. Obviously,
there is no difference between the Brans-Dicke and Einstein frames
in this case. Such particular solutions exist in the cases of one
or two internal Ricci-flat spaces. The later model is of special
interest because the internal spaces undergo the dynamical
evolution and, at the same time, the internal space total volume
is fixed. However, these models have a number of drawbacks. First,
the external space is non-accelerating and, second, these models
are unstable.

Thus in many cases, considered S-brane solutions admit stages of
the accelerating expansion of the external space. However, they
have a significant problem with the experimental bounds for the
variations of the fine structure constant.

\appendix*
\section{\label{sec:A}Brans-Dicke and Einstein frames}
\renewcommand{\theequation}{A\arabic{equation}}
\setcounter{equation}{0}

In this appendix we derive the connection between different
quantities in the Einstein and Brans-Dicke frames. Since the
result depends on the choice of the external space and both $M_0$
and $M_1$ can be the external one, we re-define by letter "E" the
external space and letter "I" the internal one, dropping the
indices 0 and 1. Further, we can perform the dimensional reduction
of action \rf{1.5} integrating over the coordinates of the
internal space\cite{RZ}:
\ba{a.1} \nonumber &S &= \frac{V_{0(I)}}{2\kappa _D^2
}\int\limits_{ \overline{M}_{(E)} } {d^{D_{(E)} }x\sqrt {\vert
\overline {g}^{(E)}\vert } }\, \, e^{d_{(I)} \beta ^{(I)}}
\biglb\{ R[\overline g^{(E)}]\\ &-&G_{II}\, \overline
g^{(E)\mu\nu}
\partial\beta^{(I)}_{\mu}\partial\beta^{(I)}_{\nu}
 + R[g^{(I)}]e^{ - 2\beta ^{(I)}}\bigrb\}\,  ,
\ea
where $V_{0(I)}$ is the constant volume of the internal space
(modulo the scale factor), $ \overline {g}^{(E)}$ is the external
spacetime metric on the manifold
$\overline{M}_{(E)}=\mathbb{R}\times {M}_{(E)}$ of the dimension
$D_{(E)}=1+d_{(E)}$, $\, G_{II}=d_{(I)}(1-d_{(I)})$, and we allow
the internal space scale factor to depend on all external
coordinates $x\in \overline{M}_{(E)}$. We also dropped scalar
field because it does not affect our results. This reduced action
is written in the Brans-Dicke frame. As next step, we remove the
explicit coupling term in \rf{a.1} by conformal transformation
\be{a.2} \overline{g}_{\mu\nu}^{(E)} = \Omega ^2\tilde
{g}_{\mu\nu}^{(E)} := \exp\biglb( - \frac{2d_{(I)} \beta
^{(I)}}{D_{(E)} - 2}\bigrb)\tilde {g}_{\mu\nu}^{(E)} \, \ee
and obtain the reduced action in the Einstein frame:
\ba{a.3} S &=& \frac{V_{0(I)}}{2\kappa _D^2 }\int\limits_{
\overline{M}_{(E)} } {d^{D_{(E)} }x\sqrt {\vert \tilde
{g}^{(E)}\vert } } \\ \nonumber &\times& \biglb\{ R[\tilde
{g}^{(E)}] - \tilde {g}^{(E)\mu\nu}
\partial\psi_{\mu}\partial\psi_{\nu} +
R[g^{(I)}]e^{2A\psi }\bigrb\}, \ea
where $\psi = - Ad_{(I)} \beta ^{(I)}$, and $A = \pm \left[(D -
2)/(d_{(I)} (D_{(E)} - 2))\right]^{1/2}$.

Thus, the metric \rf{1.1} in different gauges reads:
\ba{a.4}
 g &=& - e^{2\gamma _0 }d\tau \otimes d\tau + a_{(E)}^2 g^{(E)} + a_{(I)}^2
g^{(I)}  \nonumber \\
  &=& - dt\otimes dt + a_{(E)}^2 g^{(E)} + a_{(I)}^2 g^{(I)}  \\
 & =& \Omega ^2\left( - d\tilde {t}\otimes d\tilde {t} +
 \tilde {a}_{(E)}^2 g^{(E)} \right) +
a_{(I)}^2 g^{(I)} \nonumber \, ,
\ea
where the first line is the metric in the harmonic time gauge in
the Brans-Dicke frame, the second line is the metric in the
synchronous time gauge in the Brans-Dicke frame, and the third
line is the metric in the synchronous time gauge in the Einstein
frame. Equations \rf{a.4} show that the external scale factors in
Einstein and Brans-Dicke frames are related as follows:
\be{a.5} \tilde {a}_{(E)} = \Omega ^{ - 1}a_{(E)} \ee
and there exists the following correspondence between different
times\footnote{To have the same directions of the arrows of time,
we choose the sign plus for the square root.}:
\be{a.6}
 dt = e^{\gamma _0 (\tau )}d\tau \Longrightarrow
 t = \int {e^{\gamma _0 (\tau )}d\tau
} + \mbox{const}\, , \ee
\be{a.7} d\tilde {t} = \Omega ^{ -
1}e^{\gamma _0 (\tau )}d\tau
 \Longrightarrow \tilde {t} = \int {\Omega ^{ -
1}e^{\gamma _0 (\tau )}d\tau } + \mbox{const}\, . \ee
Additionally, it is worth of noting that Eq. \rf{a.3} explicitly
indicates the possibility of the external space acceleration (in
the Einstein frame)  in the case of the hyperbolic
compactification. The fact is that an effective potential $U_{eff}
:= -(1/2)R[g^{(I)}]\exp (2A\psi )$ is positive for $R[g^{(I)}]
<0$. Similar to the positive cosmological constant, such positive
effective potentials can result in the accelerating stages of the
Universe.

\end{document}